# A feasibility study of multi-electrode high-purity germanium detector for $^{76}$Ge neutrinoless double beta decay searching

**Yang Jingzhe** [a, b] **Tian Yang** [a, b] **Dai Wenhan** [a, b] **Yang Mingxin** [a, b] **Jiang Lin** [a, b] **Wen Jinjun** [a, b] **Xue Tao** [a, b] **Zeng Ming**[a, b] **Li YuLan** [a, b] **Zeng Zhi** [a, b] *

[a] *Department of Engineering Physics, Tsinghua University,*
  *Beijing 100084, China*
[b] *Key Laboratory of particle and radiation imaging, Ministry of education,*
  *Beijing 100084, China*
  *E-mail*: zengzhi@tsinghua.edu.cn

ABSTRACT: Experiments to search for neutrinoless double-beta (0νββ) decay of $^{76}$Ge using a high-purity germanium (HPGe) detector rely heavily on background suppression technologies to enhance their sensitivities. In this work, we proposed a pulse-shape analysis method based on a neural network (NN) and a light gradient boosting machine (lightGBM; LGB) to discriminate single-electron (background) and double-electrons (0νββ signal) events in a multi-electrode HPGe detector. In this paper, we describe a multi-electrode HPGe detector system, a data-processing system, and pulse-shape simulation procedures. We built a fully connected (FC) neural network and an LGB model to classify the single- and double-electron events. The FC network is trained with simulated single- and double-electron-induced pulses and tested in an independent dataset generated by the pulse-shape simulation. The discrimination efficiency of the FC neural network in the test set for the 0νββ double-electron events signal was 77.4%, the precision was 57.7%, and the training time was 430 min. The discrimination efficiency of LGB model was 73.1%, the precision was 64.0%, and the training time was 1.5 min. This study demonstrated that it is feasible to realize single- and double-electron discrimination on multi-electrode HPGe detectors using an FC neural network and LGB model. These results can be used as a reference for future $^{76}$Ge 0νββ experiments.

# Contents

## 1. Introduction

In the past two decades, atmospheric and solar neutrino oscillation experiments have provided evidence for the existence of nonzero mass neutrinos[1]. If the neutrino is its antiparticle, it can obtain its mass through the Majorana mass term, which will significantly advance the understanding of the nature of the neutrino[2]. The Majorana nature of neutrinos leads to lepton number violation and emerges in many standard model theories. Searching for neutrinoless double-beta (0νββ) decay is considered to be a promising method for proving the Majorana properties of neutrinos[3]. The search for 0νββ decays is also expected to give the mass ordering and absolute mass scaling of neutrinos. To investigate the 0νββ decay process, a variety of 0νββ experimental schemes have been proposed globally. Different 0νββ decay target isotopes have been proposed and used in a different experiments, including $^{76}$Ge radionuclide as the target in GERDA[4], MAJORANA[5], and CDEX[6] experiments; $^{130}$Te radionuclide as the target isotope in CUREO[7] and SNO+[8] experiments; $^{136}$Xe radionuclide as the target isotope in EXO[9], KamLAND-Zen[10], NEXT[11], and PandaX[12] experiments; $^{82}$Se as the primary target isotope in SuperNEMO[13]; and $^{116}$Cd as target isotope in a COBRA[14] experiment.

The neutrinoless 0νββ experiment is a typical extremely low-background physics experiment. It is necessary to suppress the background to an extremely low level to detect extremely rare signal events. One of the critical technologies of active background suppression is to analyze the characteristics of the waveform of the signal and discriminate the signal from the background. In 0νββ experiments, one potential background for the 0νββ signal event is a single-electron event caused by β decay of cosmogenic nuclides inside the detector and a secondary electron generated by γ-ray interaction with the detector. The single- and double-electron signals are quite similar, and the track morphology is complex, so it is challenging to distinguish them[15-17]. Therefore, developing single- and double-electronic event discrimination methods based on signal characteristics has always been a popular research topic.

High-purity germanium (HPGe) is ideal for detecting 0νββ decay because of its high energy resolution, low background, and high detection efficiency. The discrimination of single and double electrons has been seldomly investigated using a traditional HPGe detector (such as BEGe detector) because their position sensitivity is not good. The GERDA collaboration[18] and the Majorana collaboration[19] developed waveform discrimination methods, such as A and E, to discriminate single and multipoint events, but limited by the detector type (single electrode, signal readout), single- and double-electronic events cannot be discriminated.

In recent years, multi-electrode HPGe detectors have been widely used in many fields because of their position sensitivity and good energy resolution. The two main types of multi-electrode HPGe detectors by electrode structure are segmented multi-electrode HPGe and strip multi-electrode HPGe. The AGATA experiment [20] used a segmented multi-electrode HPGe detector to perform track reconstruction research to achieve measurement of the nuclear structure. The COSI experiment [21] used a strip multi-electrode HPGe detector for astronomical observations. No relevant research has been conducted, however, on the single- or double-electron event discrimination of $^{76}$Ge 0νββ.

In this study, we built experimental and simulation platforms for a strip multi-electrode HPGe detector. We examined the generation mechanism of single- and double-electron event waveform in a strip multi-electrode detector. We also studied the single- and double-electron event waveform discrimination method based on track features, fully connected (FC) neural network model, and light gradient boosting machine (lightGBM; the LGB model) and its application in $^{76}$Ge 0νββ decay experiment.

## 2. Method

### 2.1 Multi-electrode high-purity germanium detector experimental system

In this study, we used a multi-electrode HPGe detector developed by Tsinghua University[22]. We read the waveform signals of different electrodes by the evaporation of amorphous germanium on the surface of the HPGe crystal. **Figure 1** shows the structure diagram of the crystal of the detector. The germanium crystal had a diameter of 30.0 mm and a height of 10.0 mm. Seven strips of p+ Al electrodes were located on the upper surface of the germanium crystal as signal readout electrodes with a width of 2.5 mm and a center distance of 0.5 mm. To reduce the influence of surface leakage current, we installed a 2-mm-wide protective ring outside the readout electrode. The bottom lithium electrode biased the detector, and the depletion voltage of the detector was +290 V. Considering the influence of surface leakage current, the recommended operating voltage was +300 V.

As shown in figure 1, seven readout electrodes, a guard ring electrode (GR), and lithium electrodes were connected to a self-made nine-channel charge-sensitive preamplifier (CSP), and the output of each channel (except the GR channel) was connected to a reverse CR amplifier with zero-pole cancellation function. The CR output signal was then recorded and collected by a CAEN V1724 (14-bit) analog-to-digital converter (ADC) with a 100 MHz sampling rate. In the experiment, we collected 4000 samples (40 μs) for each waveform.

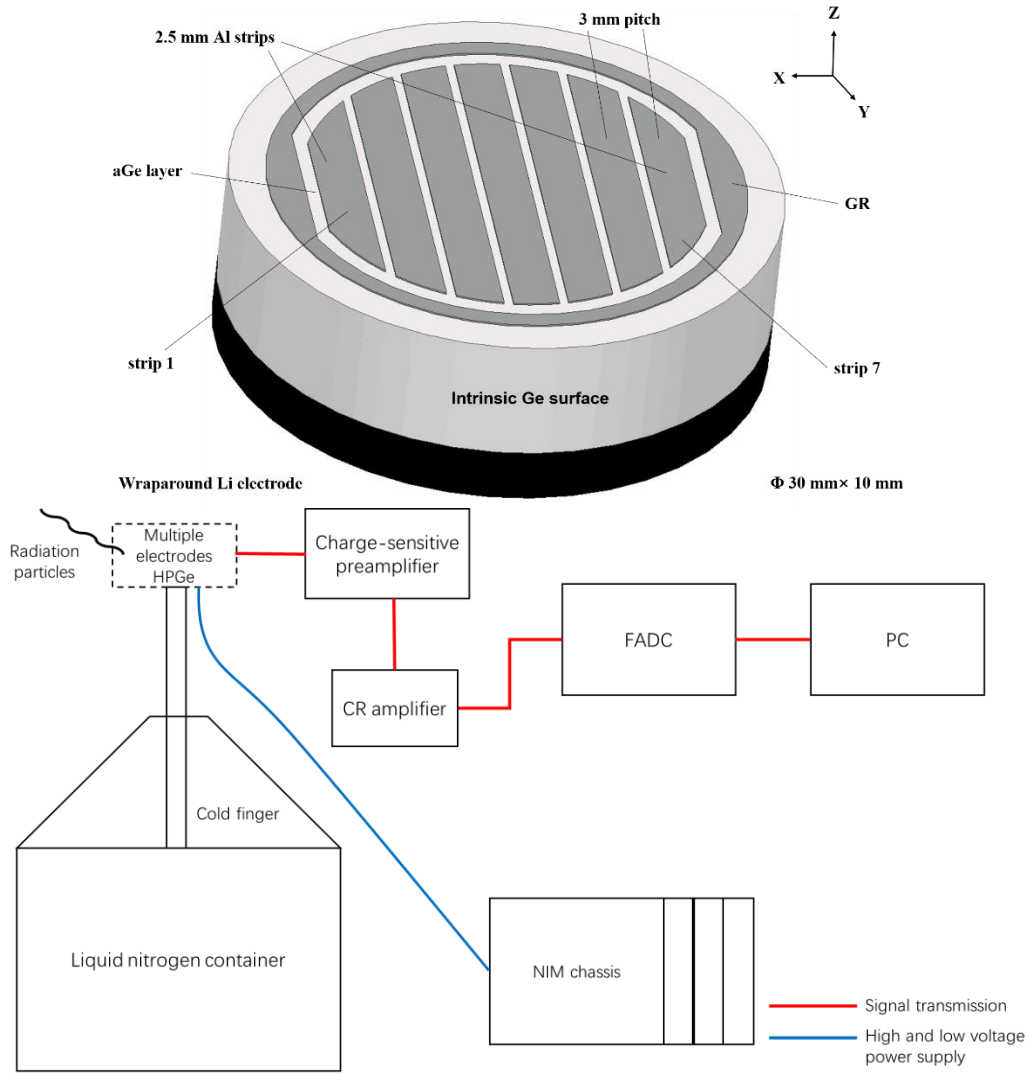

**Figure 1.** (Top image) Electrode distribution structure of the multi-electrode HPGe crystal. The inner diameter of the protective ring is 26 mm, and seven strip-shaped induction electrodes are evenly distributed in a circle with a diameter of 24 mm. (Bottom image) Diagram of the data acquisition system.

## 2.2 Multi-electrode high-purity germanium detector waveform simulation platform

In this study, we established a multi-electrode waveform simulation method for the self-developed multi-electrode HPGe detector, which provided support for detector performance optimization and subsequent waveform discrimination algorithm. We divided the pulse waveform simulation process of the HPGe detector into two parts: (1) the electric field calculation inside the germanium crystal and (2) the generation of the induced waveform caused by the drift of charge carriers. The specific process follows:

1. The electric field was calculated within the open-source SolidStateDector.jl software [23] which was capable of computing the electric and weight potential field for the multi-electrode HPGe detector.

2. The energy depositions of single and double electrons in the Ge crystal were calculated by Monte Carlo simulations using a GEANT4[24] based simulation toolkit SAGE[25]. The energy deposition value and position of each step were clustered according to their energy deposition positions: steps with energy deposition positions within 0.1 mm were clustered into an interaction point.

3. For each interaction point obtained in step 1, the drift path of induced charge carriers was calculated in a 4 ns time step. The electric field-dependent charge carrier drifts velocity was calculated using a mobility model [_ENREF_2626]: a linear relation between velocity and electric field at low field and a saturated velocity at high field.

4. The induced charge signal of each interaction point on one electrode was calculated according to the Shockley-Ramo theory [_ENREF_2727]. The charge signal of the interaction point ($Q_i$) is given by formula (1), where q is the charge of the interaction point, $\Phi_i$ is the weight potential field, $r_e(t)$ and $r_h(t)$ are the position of charge carrier electron and hold at time t respectively.

$$Q_i(t)=q[\Phi_i(\boldsymbol{r}_e(t)) - \Phi_i(\boldsymbol{r}_h(t))] \tag{1}$$

5. The output signal of an event was the sum of the signal of all interaction points within the event. **Figure 2** shows the flow chart of the waveform simulation.

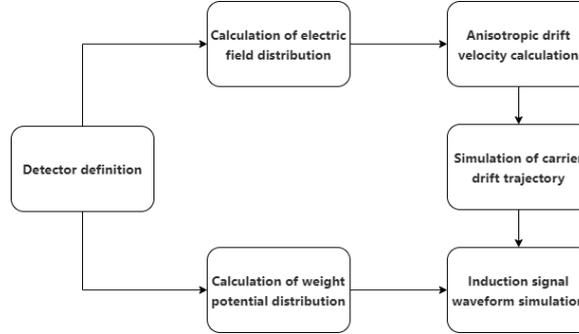

**Figure 2.** Flow chart of waveform simulation.

**2.3 Digital waveform processing system**

We established a digital waveform processing system (DPP) to extract the relevant physical characteristics from the simulated and measured waveforms. **Figure 3** shows the flow chart of the DPP module used to process the original waveform, which had four steps:

1. Data quality check. We rejected noise events using a method similar to the GERDA experiment[26]. First, waveform parameters were extracted, including baseline mean, baseline slope, baseline root mean square (RMS), and trigger position. The waveform was preliminarily discriminated according to the baseline mean and baseline slope distribution. Then, to process the waveform "stacking decision,", we excluded the same waveform containing multiple physical events and, finally, we recovered the waveform baseline. We extracted the baseline parameters using data points within the first 5 μs of the waveform, the mean value of the first 5 μs of the waveform as the baseline mean, and the linear fitting slope of the first 5 μs of the waveform as the baseline slope. A Gaussian function fit the distributions of baseline mean and baseline slope. We excluded the waveforms with a baseline mean and slope fall out of the ± 3σ range of Gaussian distribution.

2. Generation of the current waveform. We obtained current waveform by smoothing and differential processing of the charge waveform.

3. Energy reconstruction. In this part, the charge waveform transformed the exponential decay waveform into the shape of an isosceles trapezoid while ensuring that the maximum amplitude did not change. The amplitude of the charge waveform was extracted using a trapezoidal filter[27], which converted the charge waveform to a trapezoid pulse with the trapezoidal height representing the amplitude.

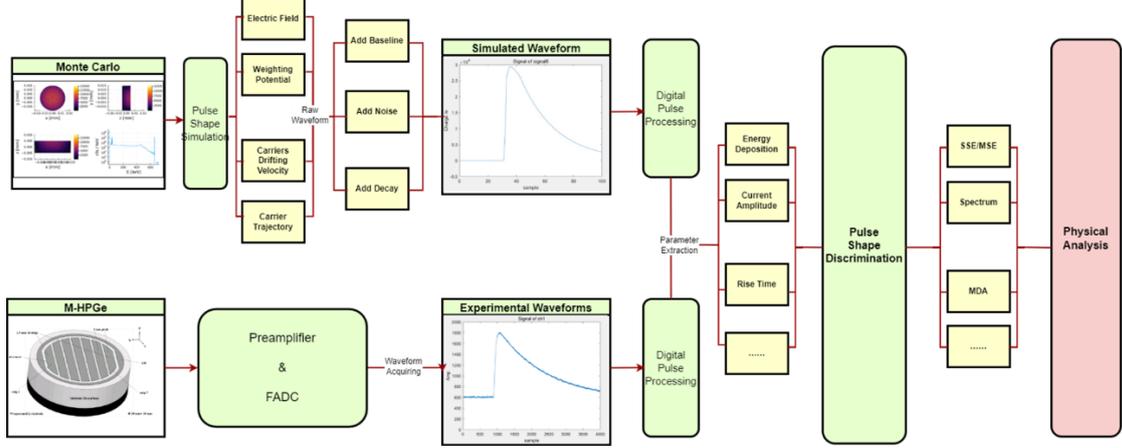

**Figure 3.** Digital waveform processing method diagram.

**2.4 Fully connected neural network structure and training process**

The neural networks currently used in particle physics are trained primarily in supervised learning[28,29]. Lots of the neural networks used in particle physics are trained by simulated data, which is the basic idea of the neural network-based single- and double-electron discrimination method in this study. We built a fully connected neural network model[30] to train single- and double-electronic event waveforms and verified the correctness of the test-set classification. The network training used the simulated waveform signal of the multi-electrode HPGe detector as the input. In the simulation, we used the $^{76}$Ge $0\nu\beta\beta$ double-electron event as the signal and used the 2039 keV single-electron event as the background.

The waveform generated by the signal and the background event was directly output in the simulation. The waveform sampling rate was selected as 250 MHz, which was the same as that used in the experiment. We extracted the input pulse of the neural network from the raw charge pulse: a ±25 point window centered at the pulse central point M (defined as the point charge pulse rises to 50% of its amplitude) was used to select the input charge pulse. As shown in **Figure 4**, these 50 points included the entire waveform rising edge area, and its amplitude would be differentiated and used as the input feature of the network. Seven different readout electrodes analog waveforms of each physical event were taken as a group. All groups of waveforms were labeled 1 for the $0\nu\beta\beta$ double-electron signal and 0 for the background. Finally, the network's output was the label prediction results that corresponded to the waveform.

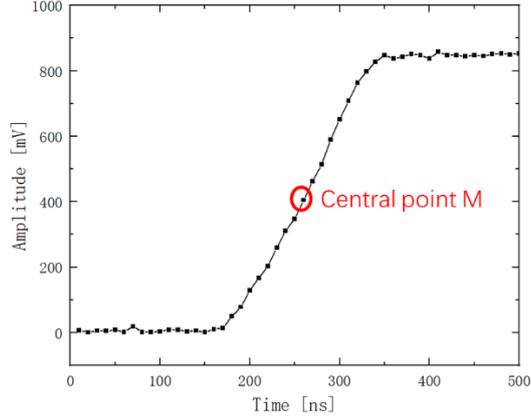

**Figure 4.** Sample of analog signal waveform of multi-electrode HPGe detector for training.

We randomly disrupted the simulated 90,000 [76]Ge 0νββ signal events and 90,000 2039 keV background events, totaling 180,000 waveform data sets. We divided the data into training, validation, and test sets according to the 70%, 20%, and 10% ratios. We used the training set for neural network training and the validation set to select the model and serve as a criterion for hyperparameter adjustment. We used the test set to evaluate network performance after training.

In this study, we implemented the neural network framework by Pytorch[31]. The structure diagram of the neural network is shown in **Figure 5**. The network's input was the multi-electrode HPGe detector seven-electrode waveform rising edge, which was evenly distributed on the 50 points $x_1$-$x_{50}$. The network extracted the waveforms of the seven electrodes separately, with three layers. The number of nodes in each layer was 32, 16, and 4. The excitation function for each layer was ReLU, and Dropout was set to retain 80% of the network. After feature extraction of seven electrode waveforms, we connected the output to a three-layer fully connected network with dimensions 16, 4, and 2. The excitation function was ReLU, and the Dropout was 50%. Finally, we conducted the two-class prediction of single- and double-electronic events.

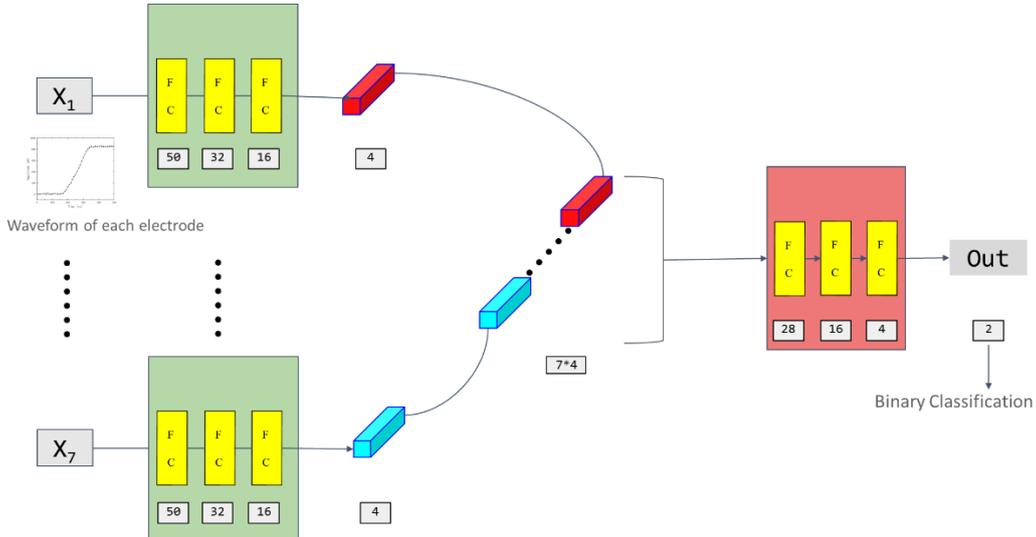

**Figure 5.** Structure diagram of fully connected neural networks for single- and double-electron event detection of [76]Ge.

The initial learning rate of single- and double-electronic discrimination network training was set to 0.001. The learning rate gradually decreased with an increase in the number of iterations in the training process. The optimizer selected the Adam model[32], which usually performed better training, and the loss function determined the cross-entropy loss function commonly used in classification problems. After multiple debugging, the training epoch was 150. Each batch size was set to 128.

**2.5 LGB structure and training process**

In addition to the FC neural network method, the decision tree is another common method used for binary classification in machine learning. Gradient boosting decision tree (GBDT) is an enduring model in machine learning. It uses the weak classifier (decision tree) iterative training to obtain the optimal model. The LGB framework is used to implement the GBDT algorithm, which supports efficient parallel training. It offers the advantages of faster training speed, lower memory consumption, better accuracy, and distributed processing of massive data[33].

In this study, we used the LGB model to classify and predict the single- and double-electron waveform discrimination of a multi-electrode HPGe detector $^{76}$Ge. The LGB framework was implemented by sklearn[34]. The input of the decision tree was 50 points $x_1$-$x_{50}$ that were distributed evenly along the rising edge of the waveform of the seven electrodes of the multi-electrode HPGe detector. We used parameter grid search through GridSearchCV[35] to determine the decision tree parameters. The final use learning type was GBDT, the learning rate was 0.008, the tree depth was 30, the number of leaves was 97, and the proportion of random sampling of features when building a weak learner was 1. The minimum number of leaf node samples was 32, and the maximum number of iterations was 2000.

**2.6 Evaluation indeX**

For the 0νββ discrimination of $^{76}$Ge, we selected common parameters to evaluate the performance of binary classifiers: accuracy, precision, recall, and area under the receiver operating characteristic (ROC) curve (AUC). For the binary classification model, we identified four combinations of predicted and actual situations (table 1).

**Table 1.** Summary of binary classification results

|  | Positive prediction | Negative prediction |
| --- | --- | --- |
| Positive truth | TP: True Positive | FN: False Negative |
| Negative truth | FP: False Positive | TN: True Negative |

Accuracy: Accuracy = (TP + TN) / (TP + FN + FP + TN). Accuracy is the percentage of the predicted correct results in the total sample. If the sample was not balanced, the accuracy would fail. Therefore, when the proportion of different categories of samples was very uneven, the category with a large proportion often became the most important factor affecting the accuracy. For the binary classification process of single- and double-electronic discrimination, accuracy was used only as a reference for the network training effect.

Recall: Recall = TP / (TP + FN). Recall is the number of correctly predicted positive cases divided by the total actual positive cases. In the actual positive record, the predicted number was positive. We used the proportion correctly predicted in all positive examples to evaluate the detection coverage of the detector for all targets to be detected. The recall represented the proportion of the useful part of the entire test result in the useful part of the data set. Recall was the discrimination efficiency for the binary classification process of single- and double-electronic discrimination.

Precision: Precision = TP / (TP + FP). Precision is the number of correctly predicted positive cases divided by the total number of predicted positive cases. In the predicted positive record, the actual number was positive. We used the proportion of true positives in the predicted results to evaluate the accuracy of the detector based on successful detection. For the binary classification process of single- and double-electronic discrimination, precision represented the proportion of the useful part in the whole detection result.

Area under the ROC curve (AUC): When drawing the ROC curve, the horizontal and vertical axes are 0 to 1, forming a square. AUC is the area enclosed by the ROC curve in this square. For the binary classification process of single- and double-electronic discrimination, AUC = 0.5 represented a random classifier; AUC < 0.5 indicated that it was worse than the random classifier and had no modeling value; and 0.5 < AUC < 1 was better than the random classifier.

## 3. Result and Discussion

We verified the pulse-shape simulation by comparing the rise time of simulated and experimental waveforms. The 10–90% rise time distribution of the $^{57}$Co source (30 cm above the germanium crystal) measured in the experiment is shown in **Figure 6**. The response functions of the CSP and CR circuits were considered in the simulation process. By comparison, the simulation waveforms well represented the measurement waveforms.

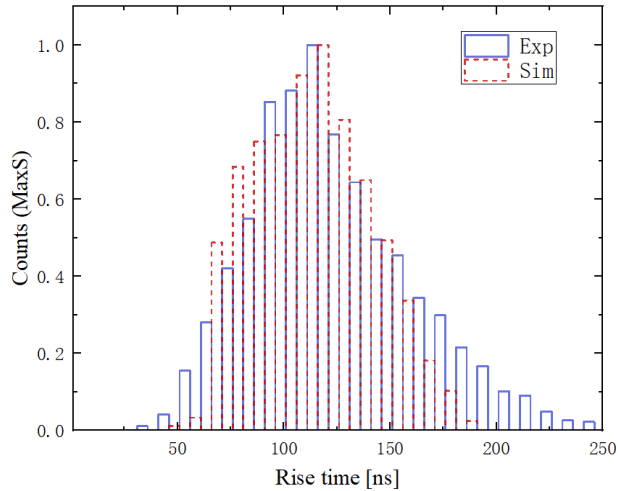

**Figure 6.** Simulation and experimental waveform rise time distribution.

### 3.1 0νββ discrimination method based on track feature

**Figure 1** shows the coordinate system for the strip multi-electrode HPGe detector used in this study. In the following simulation, the origin of the coordinate system was located at the bottom center. The position resolution of the detector for a single-site event was calculated by the signal amplitude induced by the charge collection electrode and its adjacent electrode. The position of the event in the X direction for any electrode A can be calculated by formula (2) [36]. For a given electrode structure, the spatial resolution at different positions can be calculated by the error transfer formula of formula (3). The root means the square of noise is two mV. The simulation used a 122 keV collimated gamma point source, and the Y and Z directions were fixed at the center. The simulation results of spatial resolution are shown in **Figure 7**. The values of $(S_R - S_L) / (S_R + S_L)$ at different X directions were linearly fitted, the K value was 1.78, and the fitting error was 0.026. Therefore, we ignored the error term of the K value. The resolution of the position directly below the electrode was about 0.3 mm, and the position at the edge of the electrode was about 0.15 mm. The results showed that this structure's multi-electrode HPGe detector had good position sensitivity in the X direction.

$$X = K \cdot \frac{S_R - S_L}{S_R + S_L} + X_A \tag{2}$$

$$\sigma_X^2 = \left(\frac{\partial X}{\partial S_R}\right)^2 \cdot \sigma_{S_R}^2 + \left(\frac{\partial X}{\partial S_L}\right)^2 \cdot \sigma_{S_L}^2 + \left(\frac{\partial X}{\partial K}\right)^2 \cdot \sigma_K^2 \tag{3}$$

$$= K^2 \cdot \sigma_{S_R}^2 \cdot \left[\frac{S_R + S_L - S_R + S_L}{(S_R + S_L)^2}\right]^2 + K^2 \cdot \sigma_{S_L}^2 \cdot \left[\frac{-S_R - S_L - S_R + S_L}{(S_R + S_L)^2}\right]^2 + \sigma_K^2 \cdot \left[\frac{S_R - S_L}{S_R + S_L}\right]^2$$

$$= K^2 \cdot \sigma_S^2 \cdot \frac{4S_R^2 + 4S_L^2}{(S_R + S_L)^4} + \sigma_K^2 \cdot \left[\frac{S_R - S_L}{S_R + S_L}\right]^2,$$

where $S_L$ and $S_R$ are the maximum amplitudes of the induced signals of adjacent electrodes on the left and right sides of electrode A, respectively; $X_A$ is the center position of electrode A; and K is the slope constant obtained by fitting the linear relationship between the X position and $(S_R - S_L) / (S_R + S_L)$.

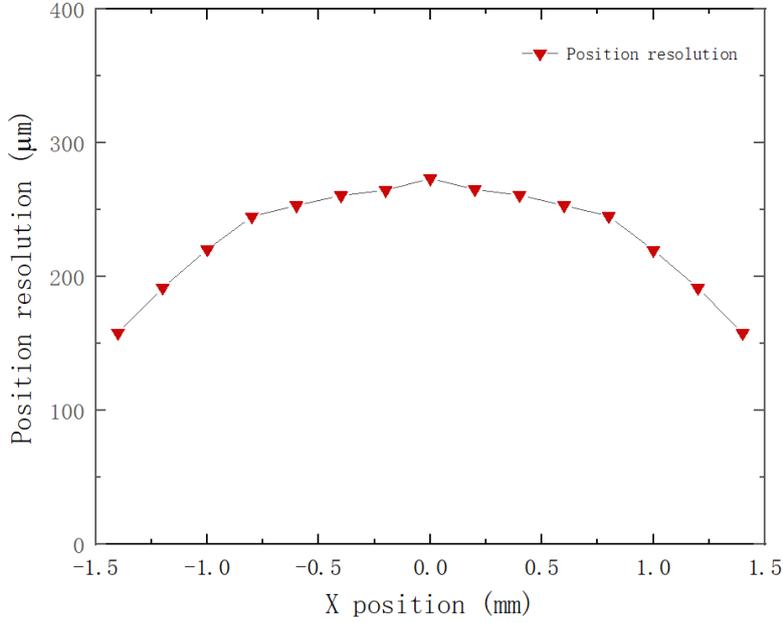

**Figure 7.** Y-direction spatial resolution comparison.

We studied the energy deposition features of the electrons in Ge crystal by establishing a concentric germanium sphere model with a maximum radius of 100 mm using the MC simulation module. Electrons of specific energy were emitted isotropically at the center of the model, and we selected the events in which all energy was deposited in the model. In this part of the event, we counted the probability of electrons depositing all energy in each germanium sphere region and established a probability model. In the simulation, we investigated 2039 keV electrons. When considering the bremsstrahlung effect, the generated secondary γ-ray had a strong penetrating ability, which led to a somewhat-larger energy deposition area of the current event. The result showed that for the electrons below 2039 keV, the energy deposition in germanium mainly occurred in the spherical region with a radius of 0.5–2 mm. This showed that although the full-energy deposition of electrons occurred in a minimal area of millimeters, if the HPGe detector had sufficient position resolution, it would be theoretically possible to identify single- and double-electron events.

To simulate the behavior of single electron events and 0νββ events, we used SAGE for simulation, using the G4EmLivermorePhysics in the low-energy electromagnetic physics list, and set a cut value of 0.1 μm for the electron. In HPGe, the interested 0νββ element was $^{76}$Ge with a maximum Q value of 2039keV. Single- and double-electron events were uniformly sampled in the HPGe crystal, and emitted electrons were generated, discriminating for events that retained all energy deposition in the HPGe. In the 0νββ decay of $^{76}$Ge, two electrons were simultaneously emitted with a total energy of 2039 keV. We simultaneously simulated single-electron events to identify their characteristic differences from 0νββ events. A single electron was emitted from the center of HPGe in a random direction. The initial energy of the electron was 2039 keV, which corresponded to the Q value of $^{76}$Ge. In the HPGe detector system, the single-electron background may have come from the β decay of the Ge cosmogenic nuclides $^{68}$Ga and $^{60}$Co, from the secondary electrons generated by the photoelectric effect, or from the single Compton scattering of photons with an energy greater than 2039 keV in the crystal.

For the two kinds of single-electron events and 0νββ events generated by GEANT4, we recorded the energy deposition and x, y, and z coordinates of each step level in the data set. The primary method to observe the 0νββ decay was to measure the total energy distribution of two electrons and find the 0νββ peak. According to the 0νββ decay model, the energy of nuclear recoil was negligible, and two electrons carried almost all of the decay energy Q; a single electron had a single-energy deposition higher end, and the energy deposition of the two ends of the double electron was relatively large[15]. In this study, we calculated the shortest path length between each pair of candidate points by looking for the higher energy deposition points in the 0νββ track of $^{76}$Ge as the endpoint candidate points and selected the candidate points with the most extended path length as the track endpoints. The distance between the single- and double-electron endpoints is shown in **Figure 8**. The waveform difference between single- and double-electronic events was negligible. In this study, the multi-electrode HPGe detector could not directly discriminate single- and double-electron events according to the track feature method. Therefore, we developed a method based on machine learning utilizing the detailed information in the signal pulse to discriminate the single- and double electron events.

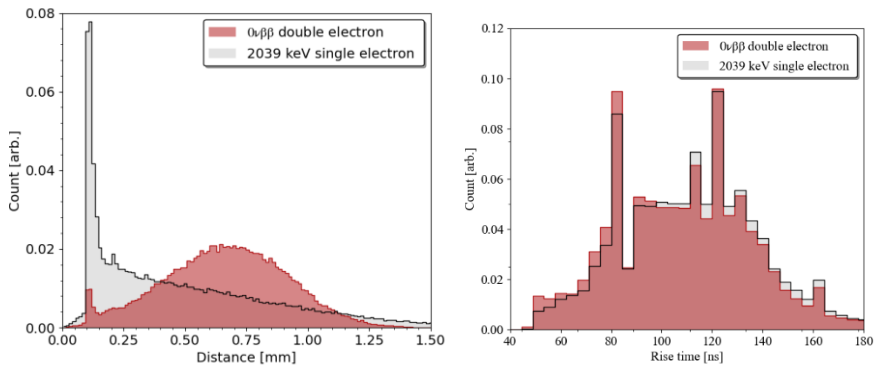

**Figure 8.** (Left) Maximum distance of the double-ended interaction point; and (Right) rise time distribution of single- and double-electron events.

### 3.2 0νββ discrimination method based on FC neural network model

We used the Tesla V100 graphics card to train the network convergence, which took about 430 minutes. **Figure 9** shows the relationship between the accuracy and loss function of the training set and the validation set with the number of training rounds after the FC network was trained. The correct rate of the training set and the validation set was about 50% at first and eventually approached 57.5% with the training. The loss function values of both were lower than 0.68. **Figure 10** shows the ROC curves to classify test sets (15,000 sets of waveforms) using the FC neural network model trained in this study. The AUC of the network was about 0.61. According to the calculation, recall was close to 77.4%, and the precision was 57.7%. The results showed that using multi-electrode high-purity detectors, the FC neural network could serve as a single- and double-electron discriminator in the search for $^{76}$Ge 0νββ decay.

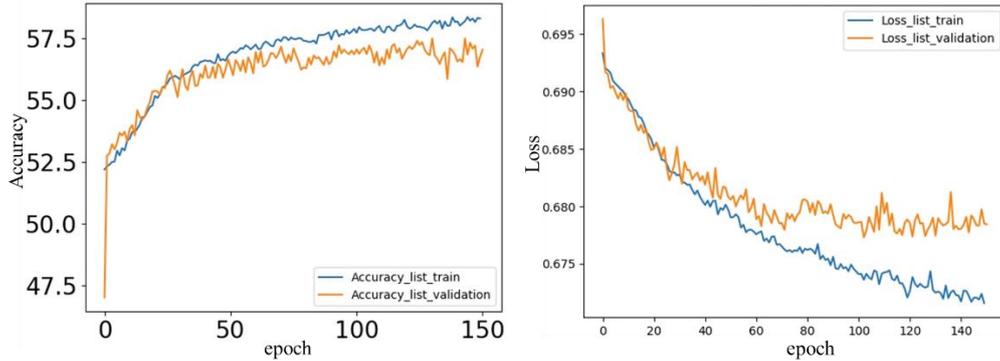

**Figure 9.** (Left) Accuracy of training set and validation set of FC neural network model; and (Right) loss of training and validation set of FC neural network model.

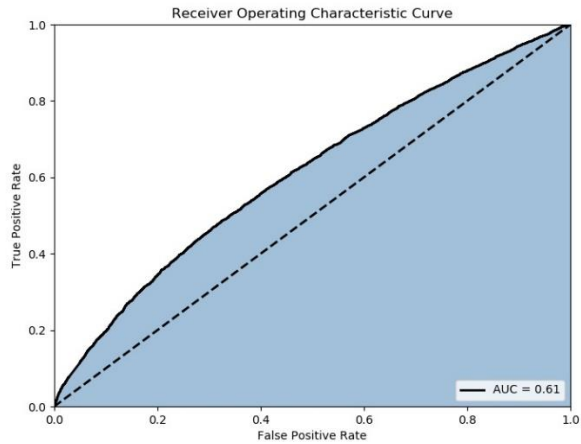

**Figure 10.** ROC curve of FC neural network model.

### 3.3 0νββ discrimination method based on LGB model

The dataset of LGB learning was the same as dataset of the FC neural network. The training of the LGB model took only 1.5 minutes. **Figure 11** shows the relationship between the accuracy and loss function of the training set and the validation set with the number of training rounds during the training process. The correct rate of the training set and the validation set was about 50% at first and eventually approached 64.3% with the training. The loss function values of both were lower than 0.64. **Figure 12** shows the ROC curves to classify the test sets (15,000 sets of waveforms) using the LGB model trained in this study. The AUC of the network was about 0.69. According to the calculated, recall was close to 73.1%, and the precision was 64.0%.

The results showed that the LGB was an effective method for searching 0νββ events of $^{76}$Ge for multi-electrode HPGe detector.

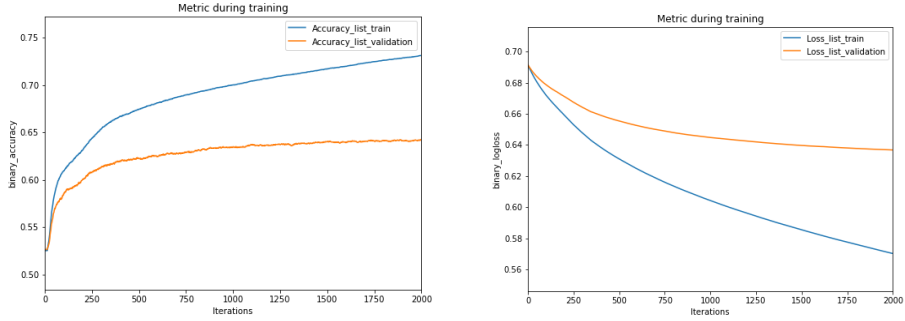

**Figure 11.** (Left) Accuracy of training set and validation set of LGB model; and (Right) loss of training and validation set of LGB model.

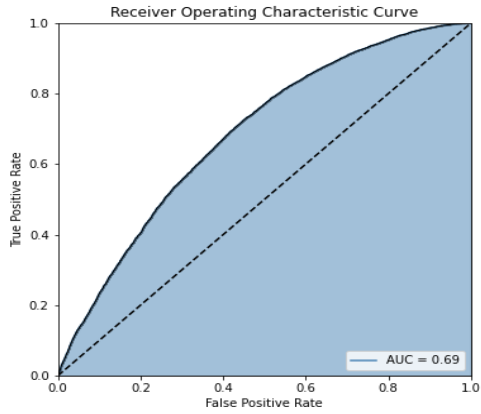

**Figure 12.** ROC curve of LGB model.

### 3.4 Effect comparison of discrimination methods

We studied the feasibility of waveform discrimination of $^{76}$Ge 0νββ single and double electrons in a multi-electrode HPGe detector according to three methods. We found that among the three methods, the track feature method was very dependent on the inherent position resolution of the detector. The detector structure in this paper could not directly and effectively utilize the characteristics of the waveform, so it was impossible to discriminate single and double electrons based on the track feature method. The FC neural network and LGB method based on machine learning could realize the extraction and learning of the waveform features, which provided a new method for multi-electrode HPGe detector for 0νββ single- and double-electron recognition of $^{76}$Ge. The AUC values of the single- and double-electron methods were 0.61 and 0.69, respectively, both greater than 0.5. The results indicated that both methods effectively discriminated the single and double electrons of $^{76}$Ge 0νββ. The metrics parameters showed that the FC-based discrimination method recall was close to 77.4%, but the precision was only 57.7%. This method took a long time, and many places still could be improved for feature extraction of FC neural networks in the follow-up work. The LGB-based discrimination method achieved 73.1% recall and 64.0% precision, and the computation time was significantly reduced.

Table 2. Effect of Three Methods on Single- and Double-Electron Event Discrimination

|  | Track feature | FC | LGB |
|---|---|---|---|
| Feasibility | FALSE | TRUE | TRUE |
| ACC | \ | 57.5% | 64.3% |
| Recall | \ | 77.4% | 73.1% |
| Precision | \ | 57.7% | 64.0% |
| AUC | \ | 0.61 | 0.69 |
| Training Time | \ | 430 min | 1.5 min |

## 4. Conclusion

In this study, we built a multi-electrode HPGe detector experimental system and waveform simulation research platform. We studied the feasibility of three methods for 0νββ single- and double-electron discrimination of $^{76}$Ge in multi-electrode HPGe detector using pulse-shape simulation. Through simulation, we found that the resolution of the position below the electrode in the vertical electrode distribution direction was about 0.3 mm, and the position of the electrode edge was about 0.15 mm. The results showed that the multi-electrode HPGe detector with this structure had good position sensitivity in the vertical electrode distribution direction. Among the three methods, the track feature method was very dependent on the inherent position resolution of the detector. The detector structure in this study could not directly and effectively utilize the characteristics of the waveform, so it was impossible to discriminate single and double electrons based on the track feature method. We built an FC neural network and LGB model of the fully connected structure to train the simulation waveforms collected by each electrode of the multi-electrode HPGe detector. The discrimination between 0νββ double-electron events and 2039 keV single-electron events was feasible. In the absence of noise interference, the discrimination efficiency of the FC neural network was 77.4%, the precision was 57.7%, and the training time was 430 min. The discrimination efficiency of the LGB model was 73.1%, the precision was 64.0%, and the training time was 1.5 min. The results showed that it was feasible to realize single- and double-electron discrimination on multi-electrode HPGe detectors using the FC neural network and LGB model. These results can be used as a reference for future $^{76}$Ge 0νββ experiments.

## Acknowledgments

This work was supported by the National Natural Science Foundation of China (Grants No. U1865205).